\documentclass{iau}
\usepackage{graphicx}

\title[JD 11.~~Silicate crystallinity of evolved stars] %% give here short title %%
{Are the silicate crystallinities of oxygen-rich evolved stars related to their mass loss rates?}

\author[Biwei Jiang, Jiaming Liu \& Aigen Li]   %% give here short author list %%
{Biwei Jiang$^1$,
 Jiaming Liu$^2$ \and Aigen Li$^3$}

\affiliation{$^1$ Department of Astronomy, Beijing Normal University,
Beijing 100875, China \\ email:{\tt bjiang@bnu.edu.cn}\\[\affilskip]
$^2$ National Astronomical Observatories, Chinese Academy of Sciences,
Beijing 100012, China \\ email:{\tt jmliu@nao.cas.cn}\\[\affilskip]
$^3$ Department of Physics and Astronomy, University of Missouri, Columbia, MO 65211, USA \\ email:{\tt lia@missouri.edu }}

\pubyear{2019}
\volume{343}  %% insert here IAU Symposium No.
\jname{Title of your IAU Symposium}
\editors{A.C. Editor, B.D. Editor \& C.E. Editor, eds.}
\begin{document}
\maketitle

\begin{abstract}
%Small amounts of pre-solar grains have survived in the matrices of primitive meteorites
%and interplanetary dust particles. Their detailed study in the laboratory with modern
%analytical tools provides highly accurate and detailed information with regard to stellar
%nucleosynthesis and evolution, grain formation in stellar atmospheres, and Galactic
%Chemical Evolution. Their survival puts constraints on conditions they were exposed
%to in the interstellar medium and in the Early Solar System.
%\keywords{Keyword1, keyword2, keyword3, etc.}
%%% add here a maximum of 10 keywords, to be taken form the file <Keywords.txt>
A sample of 28 oxygen-rich evolved stars is selected based on the presence of crystalline silicate emission features in their ISO/SWS spectra. The crystallinity, measured as the flux fraction of crystalline silicate features, is found not related to mass loss rate that is derived from fitting the spectral energy distribution.
\end{abstract}

\firstsection % if your document starts with a section,
\section{Introduction}

Crystalline silicates are identified through a series of sharp spectral lines between 10-70 micron by the ISO and Spitzer space observations (\cite[Henning 2010]{Henning2010}). They are detected in various types of objects, from the solar system objects -- comets, to pre-main sequence stars, evolved stars and distant quasars. The spectral features of crystalline silicates are detected in every stage of evolved stars: AGB stars, post-AGB stars and planetary nebulae (e.g. \cite[Jiang et al. 2013]{Jiang2013}). The degree of crystallinity, i.e. the mass percentage of crystalline silicate in all silicate dust, is found to range from a few percent to $>90\%$. What determines crystallinity has long been debated (\cite[Liu et al. 2014]{Liu2014}). Mass loss rate is thought to be an important factor. Theoretical calculations have shown that amorphous silicates cannot be crystallized in stars of low mass loss rate because the dust cannot be heated to temperatures high enough for crystallization, and that crystalline silicates can only be formed in stars undergoing substantial mass-loss with a critical value of $\dot{M} > 10^{-5} {\rm M_\odot yr^{-1}}$ and having high dust column densities
%% e.g.,
%%\cite[Draine 2003]{Draine03})
(e.g. \cite[Gail et al. 2003]{Gai1999}).  \cite[Jones et al. (2012)]{Jon2012} analysed the Spitzer/IRS spectra of 315 evolved stars and found that the mass-loss rates of the stars exhibiting the crystalline silicate features at 23, 28 and 33 $\mu$m span over 3 dex, down to $10^{-9} {\rm M_\odot yr^{-1}}$. They investigated the possible correlation between $\dot{M}$  and the silicate crystallinity by examining the relation of $\dot{M}$ with the strengths of the 23, 28 and 33 $\mu$m features measured by their equivalent widths, but found no correlation. \cite[Kemper et al. (2001)]{Kem2001} performed an extensive radiative transfer calculation of the model IR emission spectra for O-rich AGB stars of mass-loss rates ranging from $10^{-7} {\rm M_\odot yr^{-1}}$ to $10^{-4} {\rm M_{\odot} yr^{-1}}$  and of a wide range of crystallinities up to 40 percent. They also found that crystallinity is not necessarily a function of mass-loss rate.

\section{Data and method}

We selected a sample of nearby 28 oxygen-rich evolved stars (mainly AGB stars and red supergiants) which show prominent spectral features of crystalline silicate as well as amorphous silicate in their ISO spectra. The mass loss rate is calculated by fitting the spectral energy distribution from (ultraviolet-)optical to far-infrared based on the photometry in the UBVRI, 2MASS/JHK, WISE and IRAS bands. The ‘2-DUST’ radiative transfer code (\cite[Ueta \& Meixner 2003]{Uet2003})
is used to derive the mass loss rate and other stellar and circumstellar parameters.

The silicate crystallinity ($\eta_{\rm csi,f}$) is characterized with the flux ratio of the emission features of crystalline silicates to that of all (crystalline + amorphous) silicates. This measure of crystallinity differs from the  common measure of the mass ratio of crystalline to total silicate. The calculation of silicate mass suffers generally significant uncertainties from dust temperature in addition to the opacity since the species is not clearly identified. The flux ratio would represent the mass ratio if the UV/V/IR opacities of crystalline and amorphous silicates are comparable. The flux of crystalline and amorphous silicate is calculated with the PAHFIT code by decomposing the ISO spectrum into continuum and spectral features of  amorphous and crystalline silicate.

\section{Result}

\begin{figure}[h]
% \vspace*{-2.0 cm}
\begin{center}
 \includegraphics[width=3.8in]{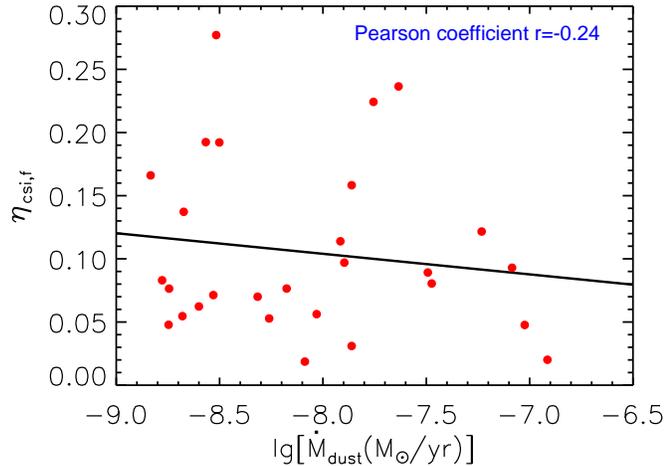}
% \vspace*{-1.0 cm}
 \caption{Relation between mass loss rates and silicate crystallinities of O-rich evolved stars.}
   \label{fig1}
\end{center}
\end{figure}

The derived mass loss rate and crystallinity have  a Pearson correlation coefficient of -0.24 (Figure \ref{fig1}), which indicates that the silicate crystallinities and the mass-loss rates of these oxygen-rich evolved stars are not correlated (\cite[Liu et al. 2017]{Liu2017}). A further check also found no relation of crystallinity with stellar luminosity or effective temperature.


\begin{thebibliography}{}
%\bibitem[Fitzpatrick \& Massa (1998)]{FM1998}
%{Fitzpatrick, E., \& Massa, D.} 1998, \textit{ApJ}, 328, 734
\bibitem[Gail \& Sedlmayr (1999)]{Gai1999}{Gail H., \& Sedlmayr, E.} 1999, \textit{A\&A}, 347, 594
\bibitem[Henning (2010)]{Henning2010}{Henning T.} 2010, \textit{ARA\&A}, 48, 21
\bibitem[Jiang et al.(2013)]{Jiang2013} {Jiang, B. W., Zhang, K.,
        Li, A., \& Lisse, C.}  2013, \textit{ApJ}, 765, 72
\bibitem[Jones et al. (2012)]{Jon2012} {Jones, O. et al.} 2012,
       \textit{MNRAS}, 427, 3209
\bibitem[Kemper et al. (2001)]{Kem2001} {Kemper, F.,
         Waters, L., de Koter, A.,
         \& Tielens, A.}  2001, \textit{A\&A}, 369, 132
\bibitem[Liu \& Jiang (2014)]{Liu2014} {Liu, J. M., \& Jiang, B. W.}
        2014, \textit{Progress in Astronomy}, 32, 2
\bibitem[Liu et al. (2017)]{Liu2017} {Liu, J. M., Jiang, B. W., Li, A.G., \& Gao, J.} 2017, \textit{MNRAS}, 466, 1963
%\bibitem[Molster et al. (2002)]{Mol2002} {Molster, F.,
%         Waters, L., Tielens, A., et al.} 2002, \textit{A\&A}, 382, 241
\bibitem[Ueta \& Meixner (2003)]{Uet2003} {Ueta, T., \& Meixner, M.}
         2003, \textit{ApJ}, 586, 1338
\end{thebibliography}
\end{document}